\documentclass{article}
\usepackage{arxiv}

\usepackage[utf8]{inputenc} % allow utf-8 input
\usepackage[T1]{fontenc}    % use 8-bit T1 fonts
\usepackage{hyperref}       % hyperlinks
\usepackage{url}            % simple URL typesetting
\usepackage{booktabs}       % professional-quality tables
\usepackage{amsfonts}       % blackboard math symbols
\usepackage{nicefrac}       % compact symbols for 1/2, etc.
\usepackage{microtype}      % microtypography
\usepackage{lipsum}
\usepackage{graphicx}
\graphicspath{ {./images/} }
\usepackage{booktabs}  % Add this in your preamble if not already included
\usepackage{array}
\usepackage{enumitem}

\usepackage[round]{natbib} 
\usepackage{xcolor}
\usepackage{caption}
\usepackage{hyperref}
\usepackage{cleveref}
\crefname{figure}{Fig.}{Figs.}
\hypersetup{
    colorlinks=true,
    linkcolor=blue,    % for \ref, \pageref, TOC
    citecolor=blue,    % for \cite
    urlcolor=blue      % for \url
}

\newcommand{\RN}[1]{\uppercase\expandafter{\romannumeral#1}}

\title{Agentic Large Language Models for Automated Structural Analysis of 3D Frame Systems}

\author{
Ziheng Geng$^{1}$,
Ian Franklin$^{1}$,
Santiago Martinez$^{2}$,
Jiachen Liu$^{3}$,
Yunhe Zhao$^{4}$,
Minghui Cheng$^{1,2\dagger}$\\
\\
$^{1}$Department of Civil and Architectural Engineering, University of Miami, Coral Gables, FL 33146, USA\\
$^{2}$School of Architecture, University of Miami, Coral Gables, FL 33146, USA\\
$^{3}$HBC Engineering Company, Doral, FL 33178, USA\\
$^{4}$Department of Electrical and Computer Engineering, University of Miami, Coral Gables, FL 33146, USA\\
\\
$^{\dagger}$Corresponding author: \texttt{minghui.cheng@miami.edu}
}

\begin{document}
\maketitle
\begin{abstract}
Large language models (LLMs) have emerged as powerful foundation models with strong reasoning capabilities across domains. Beyond reactive text generation, agentic LLMs enable autonomous workflow execution through modular task decomposition and coordinated tool use. In structural engineering, recent efforts have developed agentic LLMs for automated analysis of plane frames. However, their extension to 3D frames remains underexplored due to challenges in irregular geometric representation, topological consistency, and long-horizon reasoning. This paper proposes an agentic LLM framework for automated structural analysis of 3D frames from natural language inputs. Irregular 3D frames are represented by projection onto a 2D plan, where orthogonal gridlines define spatial coordinates and a matrix of number of stories encodes vertical extrusion of each grid cell. Building on this representation, the framework establishes a multi-agent pipeline: a problem analysis agent parses input into structured JSON; a floor decomposition agent derives the spatial layout of each floor; the 3D geometry is assembled by node, girder, slab, and column agents; support and load agents assign boundary and loading conditions, and code translation agents generate executable SAP2000 script. Evaluated on ten representative 3D frames, the proposed framework achieves an average accuracy of 90\% across repeated trials, demonstrating consistent and reliable performance.
\end{abstract}

% keywords can be removed
\begin{quote}
\end{quote}

\section{Introduction}
Large language models (LLMs), such as GPT \citep{openai2026gpt55}, Gemini \citep{deepmind2026gemini31pro}, and Claude \citep{anthropic2026claudeopus47}, are emerging as powerful foundation models for natural language understanding, reasoning, and generation. Developed through a multi-stage training pipeline involving pretraining, fine-tuning, and post-training alignment, state-of-the-art (SOTA) LLMs have demonstrated strong capabilities in instruction following \citep{zeng2024evaluating, qin2024infobench}, contextual understanding \citep{bai2024longbench, cheng2026comprehensive}, logical inference \citep{parmar2024logicbench, cheng2025empowering}, symbolic reasoning \citep{xu2024symbol, mirzadeh2025gsm}, and code generation \citep{jimenez2024swe, jain2025livecodebench}. These capabilities enable LLMs to interpret user intents and generate coherent responses across a wide range of tasks. However, vanilla LLMs primarily function as reactive text generators: they receive a prompt and produce a response. To enhance their practical utility, agentic LLMs have emerged as a transformative paradigm that extends LLMs toward autonomous systems capable of planning, reasoning, and executing complex workflows \citep{wang2024survey}. Enabled by techniques such as chain-of-thought prompting \citep{wei2022chain, wang2022self}, tool-augmented reasoning \citep{yao2022react, schick2023toolformer}, and retrieval-augmented generation \citep{borgeaud2022improving, gao2023retrieval}, these systems can decompose complex objectives into modular subtasks, orchestrate external tools, incorporate domain-specific knowledge, and complete multi-step executions, with minimal human intervention. This agentic paradigm has therefore attracted growing interest in automating scientific and engineering tasks that demand intensive manual efforts and specialized domain expertise.

Structural engineering represents one such domain where automation offers substantial practical value. In practice, engineers translate design intent into finite element (FE) models for structural analysis. Although commercial platforms such as SAP2000 \citep{sap2000} and ETABS \citep{etabs2023} provide powerful simulation environments, FE model construction remains largely manual. Engineers are required to define nodal coordinates and element connectivity to assemble the structural geometry, and then assign boundary conditions, material properties, and loading patterns to corresponding components. These operations are commonly performed through graphical user interfaces (GUIs) and involve repetitive navigation, selection, and verification. As structural systems grow in scale and geometric complexity, this modeling burden escalates significantly. Consequently, the prevailing manual workflow is time-consuming, error-prone, and difficult to scale, constituting a critical bottleneck in the structural design and analysis pipeline.

Recent studies have begun to develop agentic LLMs for automated structural design and analysis. Initial efforts have revealed notable limitations of general-purpose LLMs in conducting structural analysis \citep{wan2025som}. To addressed these limitations, \cite{liu2026large} reframed structural analysis as a code generation task and developed an LLM agent that could reliably generate OpenSeesPy scripts for beam analysis with robust generalization across varying boundary and loading conditions. As the scope expanded to 2D frame systems, spatial reasoning and long-horizon reliability emerged as critical bottlenecks. Subsequent work addressed these challenges through developing domain-specific prompts to constrain spatial reasoning \citep{liang2025integrating}, decomposing geometric assembly into stepwise plans to enhance topological consistency \citep{geng2025lightweight}, and introducing verification mechanisms to mitigate error accumulation during multi-step modeling \citep{geng2026novel}. Parallel efforts have broadened the applicability of agentic LLMs across multiple software platforms: \cite{geng2026automating} developed a two-stage pipeline for automated 2D frame analysis using OpenSees, SAP2000, and ETABS. Extending the scope from structural analysis to design, recent studies demonstrated the effectiveness of multi-agent coordination for code-compliant reinforced concrete design \citep{chen2025multi} and optimal design of ultra-high-performance concrete beams \citep{chen2026multi}. \cite{liang2025automating} introduced MASSE, a multi-agent system that replicates the structural design workflow by integrating code retrieval, structural response simulation, and safety verification. Collectively, these studies demonstrate the potential of agentic LLMs to bridge natural language problem descriptions and executable structural engineering workflows.

Despite these advancements, existing agentic LLMs for structural analysis remain largely constrained to plane beam and frame systems, which capture only a simplified representation of real-world building structures. Extending these frameworks to 3D frame systems is non-trivial and introduces three major challenges. First, irregular 3D frame geometries require a semantically unambiguous representation that LLMs can reliably interpret. These frames may involve plan asymmetry, layout variations across floors, and varying story heights, all of which must be clearly defined to avoid ambiguity during LLM reasoning. Second, topological consistency becomes more difficult to maintain. Even in 2D settings, LLMs lack a basic understanding of structural connectivity concepts such as shared nodes and elements, leading to duplicated nodes, missing members, and invalid connections \citep{geng2025lightweight}. This issue is further amplified in 3D frames, where nodes, girders, slabs, and columns must be coordinated both within each floor plan and across stories. Third, constructing a 3D structural model requires a substantially longer inference chain than its 2D counterpart. A 3D frame involves a larger set of structural components, and modeling operations such as material assignment and load application become more complex because they must be mapped to the corresponding components within this semantically dense model space. Such long-horizon reasoning steps increase the risk of hallucination and error accumulation during model generation.

This paper addresses these challenges by proposing an agentic LLM framework for automated structural analysis of 3D frame systems. First, a structured geometric representation scheme is introduced where 3D frames are projected onto a 2D plan. By utilizing orthogonal gridlines to define plan coordinates and a matrix of number of stories (MNS) to specify vertical extrusion for each grid cell, this representation enables irregular 3D geometries to be described in a concise and semantically clear form. Building on this representation, a multi-agent pipeline is developed to convert textual problem descriptions into executable structural modeling scripts. The workflow begins with a problem analysis agent that parses the input into a structured JSON format. Next, a floor decomposition agent derives the spatial layout for each floor from the MNS. Within each floor, node, girder, and slab agents operate in parallel to generate nodal coordinates and in-plane connectivity, while a column agent establishes inter-story connectivity. Support and load agents assign boundary conditions and external loads to the corresponding structural components, and code translation agents convert the assembled model information into SAP2000 scripts. The agentic LLMs are evaluated on ten representative 3D frames with various geometric configurations. Results show that the proposed framework achieves an average accuracy of 90\% across repeated trials, significantly outperforming SOTA general-purpose LLMs. It also demonstrates high computational efficiency and cost-effectiveness, with an average runtime of less than three minutes and a cost of less than USD 0.20 per run.

\section{Geometric Representation and Benchmark Design}
\label{sec:headings}

\subsection{Structured textual description template}
\begin{figure*}[htbp]
\centering
% \captionsetup{justification=centering}
\includegraphics[width=0.85\textwidth]{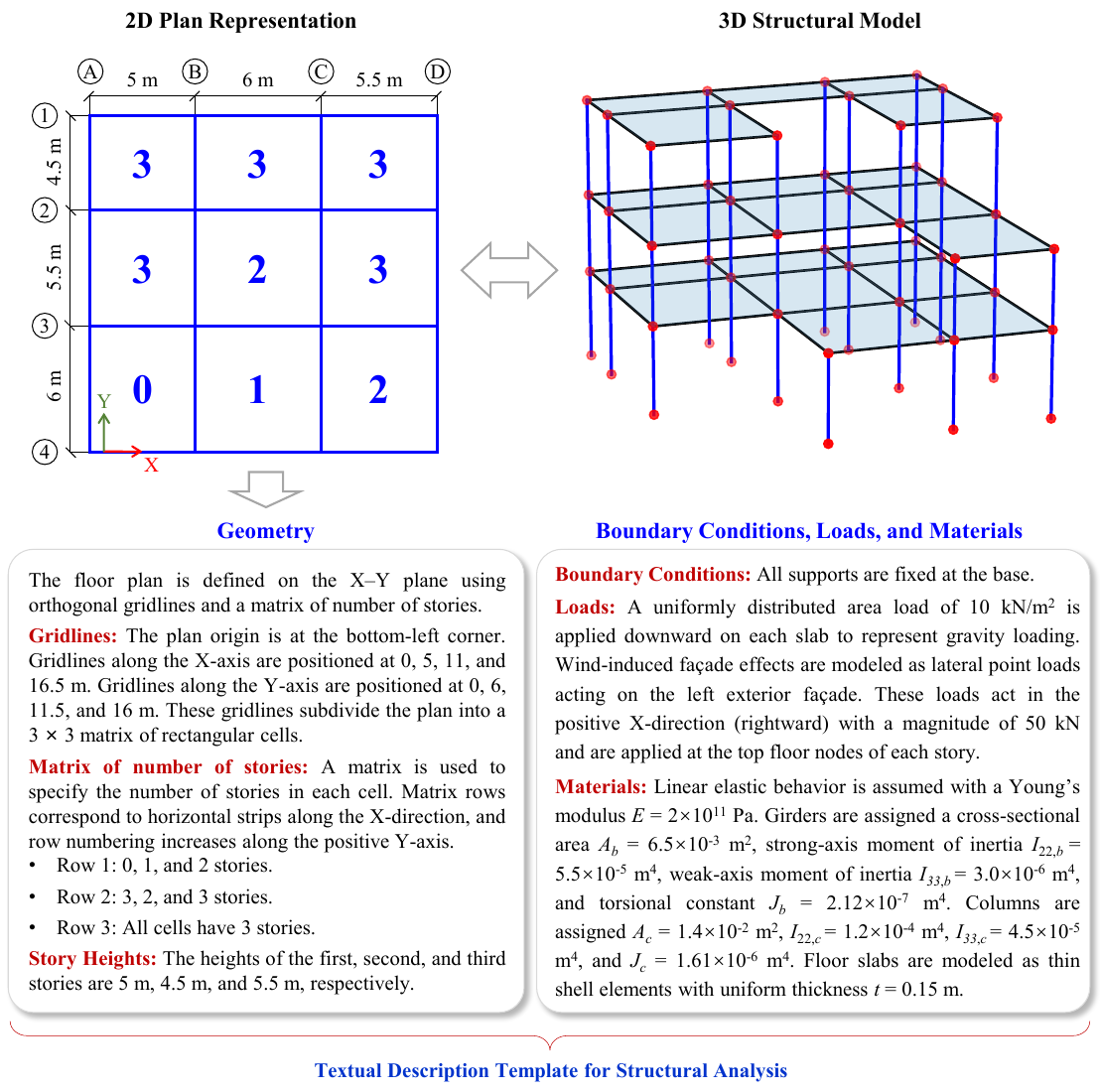}
\caption{Textual description template for automated structural analysis of 3D frame systems.}
\label{Figure1}
\end{figure*}

To enable automated structural analysis of 3D frame systems from natural language inputs, this study first establishes a structured geometric representation that can be consistently interpreted by LLMs. As illustrated in \cref{Figure1}, a 3D frame is represented by projecting its geometry onto a 2D plan. The resulting configuration is described using two components: an orthogonal gridline system and an MNS. The gridlines define the coordinate system on the X–Y plane, and their intersections and enclosed rectangular cells provide a consistent reference for locating structural components, including columns, girders, and slabs. Building on this plan-level layout, the MNS assigns a scalar integer to each grid cell to specify the number of stories present at that location. For instance, a value of 0 indicates a void or an atrium, whereas a value of 3 indicates that the corresponding cell extends vertically through three stories. Together with the specified story heights, the MNS enables reconstruction of the vertical extrusion of the frame from the 2D plan. This formulation transforms complex 3D topology into a structured representation that is highly compatible with the parsing and reasoning capabilities of LLMs.

In addition to geometry, the textual description template also defines boundary, loading, and material parameters required for structural analysis, as shown in \cref{Figure1}. Specifically, boundary conditions specify the location and type of support. By default, all supports are fixed at the base. Loading conditions comprise two components: a uniformly distributed area load applied downward on each slab to represent gravity loading, and lateral point loads applied to the left exterior façade to simulate wind effects. Material properties are specified separately for girders and columns, including Young’s modulus, cross-sectional area, strong- and weak-axis moments of inertia, and torsional constant. Floor slabs are defined by element type and uniform thickness. Together, these geometric, boundary, loading, and material descriptions constitute a complete and unambiguous problem specification that serves as the natural language inputs to the proposed agentic LLMs.

\begin{figure*}[htbp]
\centering
% \captionsetup{justification=centering}
\includegraphics[width=0.85\textwidth]{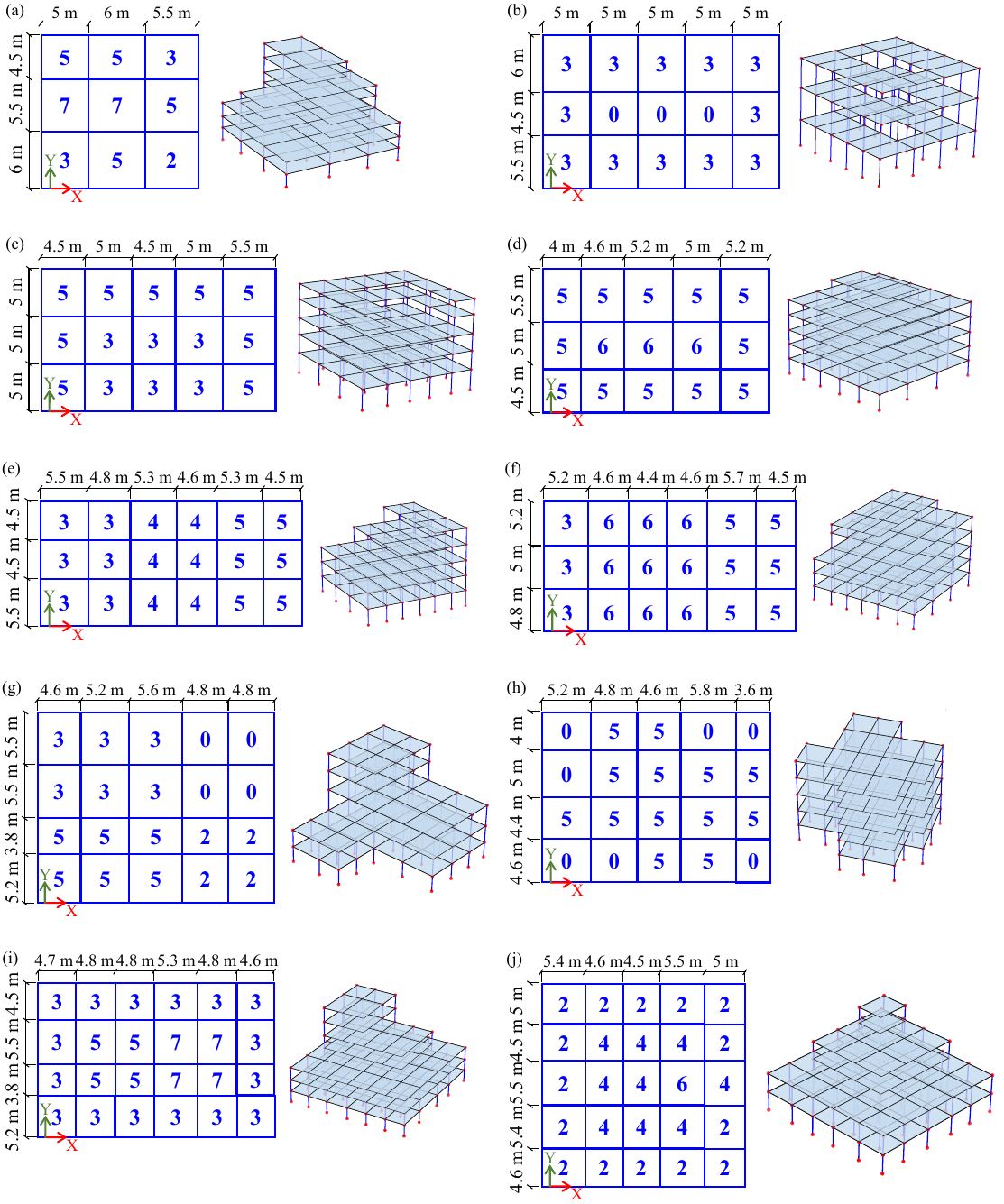}
\caption{Benchmark problems comprising ten representative 3D frame systems with irregular geometric configurations.}
\label{Figure2}
\end{figure*}

\subsection{Benchmark dataset}
To evaluate the capability of the proposed agentic LLMs in long-horizon 3D structural modeling, a benchmark dataset comprising ten representative 3D frame systems is developed, as demonstrated in \cref{Figure2}. The benchmark cases are designed to cover three major types of geometric complexity commonly encountered in engineering practice: height irregularity, plan asymmetry, and discontinuous layouts. The plan grids range from 3 × 3 to 4 × 6, with number of stories varying from 0 to 7 across grid cells. Representative configurations include stepped setbacks, internal voids, and asymmetric layouts such as L-shaped, U-shaped, and cross-shaped plans. Non-uniform grid spacing is also incorporated to assess the generalization capacity of the framework beyond regular grid systems. Collectively, these configurations constitute a comprehensive and challenging testbed for evaluating automated 3D structural modeling from natural language inputs.

For each benchmark problem, a textual description is formulated using the template introduced in \cref{Figure1}. The geometric information is defined by the corresponding gridline system and MNS shown in \cref{Figure2}, while the boundary conditions, loading patterns, and material properties are held constant across all problems and follow the illustrative setup in \cref{Figure1}. Each benchmark problem is evaluated over ten repeated trials using the same input descriptions. In each trial, the LLMs produce a SAP2000 script, which is executed to obtain structural analysis results. For reference, the authors manually construct ground truth SAP2000 models for each benchmark case. The responses from the generated model are compared with those from the ground truth model at selected key locations. A trial is classified as accurate only when the relative errors of all monitored response quantities are below 1\%. The accuracy of each benchmark problem is then calculated as the ratio of accurate trials to total trials. This protocol assesses not only the syntactic executability of the generated scripts, but also the analytical correctness of the resulting structural models.

\begin{figure*}[htbp]
\centering
% \captionsetup{justification=centering}
\includegraphics[width=0.85\textwidth]{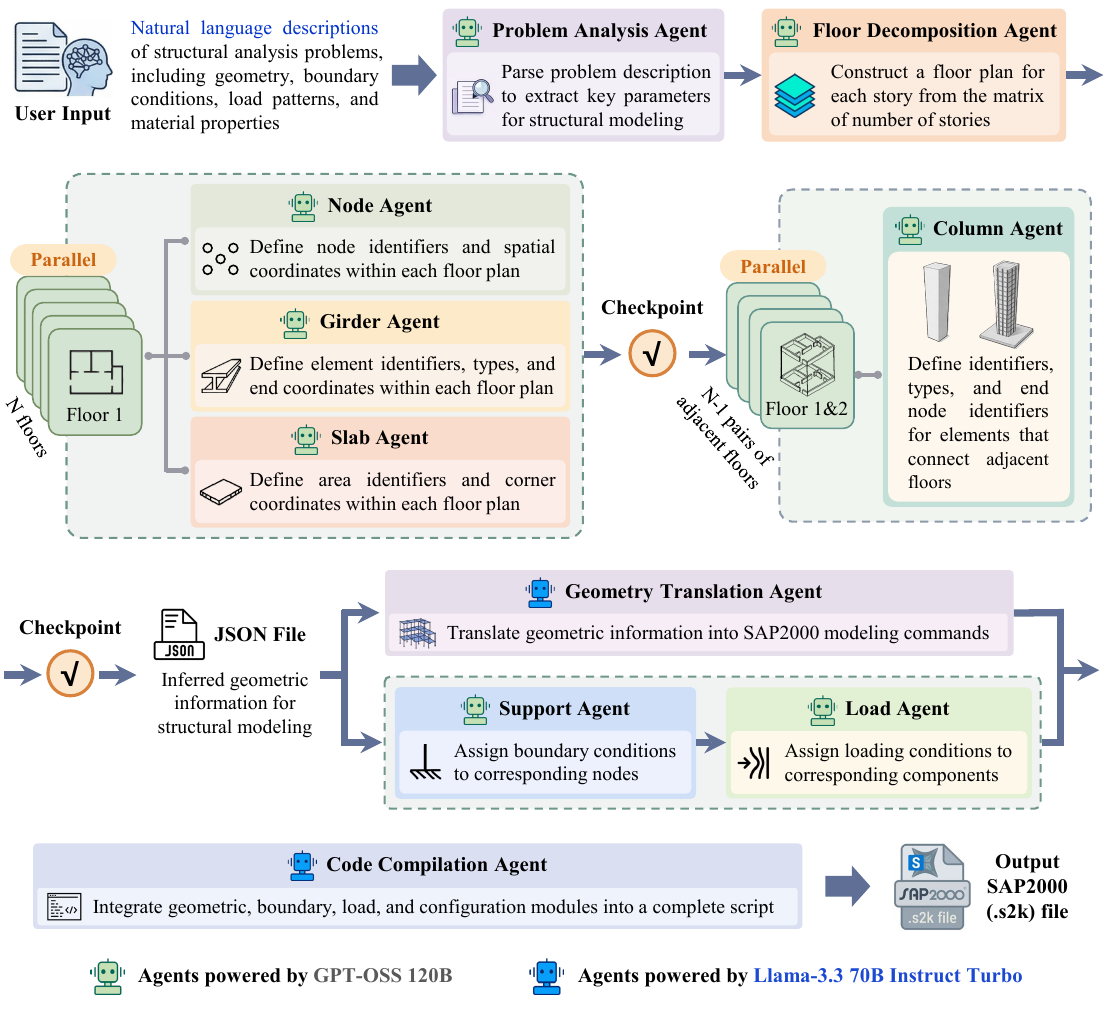}
\caption{ Multi-agent architecture of the proposed agentic LLMs for automated structural analysis.}
\label{Figure3}
\end{figure*}

\section{Agentic Large Language Models for Automated Structural Analysis}
\label{sec:others}

To address the challenges of topological consistency and long-horizon reasoning in automated 3D structural modeling, this section proposes an agentic LLM framework using multi-agent architecture. Instead of relying on a monolithic LLM to generate the complete SAP2000 script in a single inference pass, the proposed framework decomposes the modeling task into a sequence of manageable sub-tasks. Each subtask is delegated to a specialized agent operating within a clearly scoped domain, as illustrated in \cref{Figure3}. This decomposition strategy serves three purposes: (i) constrain each agent’s reasoning scope to reduce hallucination, (ii) support modular checkpoints for error detection and correction, and (iii) enable parallelization to improve computational efficiency. Intermediate structured representations, particularly JSON schemas, are introduced between agents to ensure consistent information transfer and maintain semantic coherence throughout the pipeline. The framework receives a textual problem description as input, specifying geometric configuration, boundary conditions, loading patterns, and material properties. The output is an executable SAP2000 script that defines the structural model and can be directly imported into SAP2000 for structural analysis.

\subsection{Overall multi-agent architecture}
\label{sec:overall_system}

The overall architecture of the proposed agentic LLMs consists of three sequential stages: problem interpretation, modeling information inference, and code translation. Each stage is implemented through one or more specialized agents, as illustrated in \cref{Figure3}. In the problem interpretation stage, a problem analysis agent parses the natural language inputs and extracts key parameters required for structural modeling, including gridline locations, MNS, story heights, support conditions, load patterns, and material properties. These parameters are organized into a structured JSON schema that serves as a standardized information carrier for downstream agents.

In the modeling information inference stage, a floor decomposition agent first processes the MNS to derive floor-level occupancy layouts. Specifically, the agent constructs a 2D plan for each story by comparing the story index with the matrix value assigned to each grid cell. This step converts the 3D structural topology into a series of stacked 2D floor plans. Building on these floor-level layouts, node, girder, and slab agents operate in parallel to construct nodal coordinates and in-plane connectivity, while a column agent establishes inter-story connectivity between adjacent floors. Checkpoints are introduced at key transitions to validate the consistency of node identifiers, element endpoints, slab corner nodes, and inter-story connections before information is passed downstream. Particularly, the geometric modeling pipeline, encompassing floor decomposition through column generation, is detailed in Section 3.2. The verified geometry is then passed to support and load agents, which assign boundary conditions and external loads to the corresponding structural components, as described in Section 3.3.

In the code translation stage, a geometry translation agent converts the verified structural geometry into SAP2000 modeling commands, and a code compilation agent integrates all geometric, boundary, load, and configuration modules into a complete and executable SAP2000 script. Further details of this stage are provided in Section 3.4. The framework employs two lightweight LLM backbones according to task requirements: GPT-OSS 120B is assigned to agents handling complex spatial inference due to its strong reasoning capability, whereas Llama-3.3 70B Instruct Turbo is used for code translation tasks due to its precise instruction-following capabilities. This model allocation is consistent with prior findings that different LLM backbones exhibit complementary strengths in agentic structural modeling workflows \citep{geng2026novel}.

\subsection{Geometry generation}
\label{sec:stage1}

Geometry generation is a core challenge in automated 3D structural modeling because structural components must be created with consistent topology across both horizontal floor plans and vertical inter-story connections. This process requires the coordinated generation of numerous spatially distributed components, including nodes, girders, slabs, and columns, across multiple floors and directions. Errors in any intermediate representation, such as duplicated nodes, misaligned elements, or missing slab corners, can propagate to subsequent modeling steps and compromise the final SAP2000 model. To address this challenge, the proposed framework adopts a divide-and-reconstruct strategy. The 3D frame is first decomposed into a sequence of independent floor-level generation tasks, and the global 3D topology is then reconstructed by connecting adjacent floors. As illustrated in \cref{Figure3}, the geometry generation proceeds through three sequential steps: floor decomposition, floor-level component generation, and inter-story column generation.

\begin{figure*}[htbp]
\centering
% \captionsetup{justification=centering}
\includegraphics[width=0.85\textwidth]{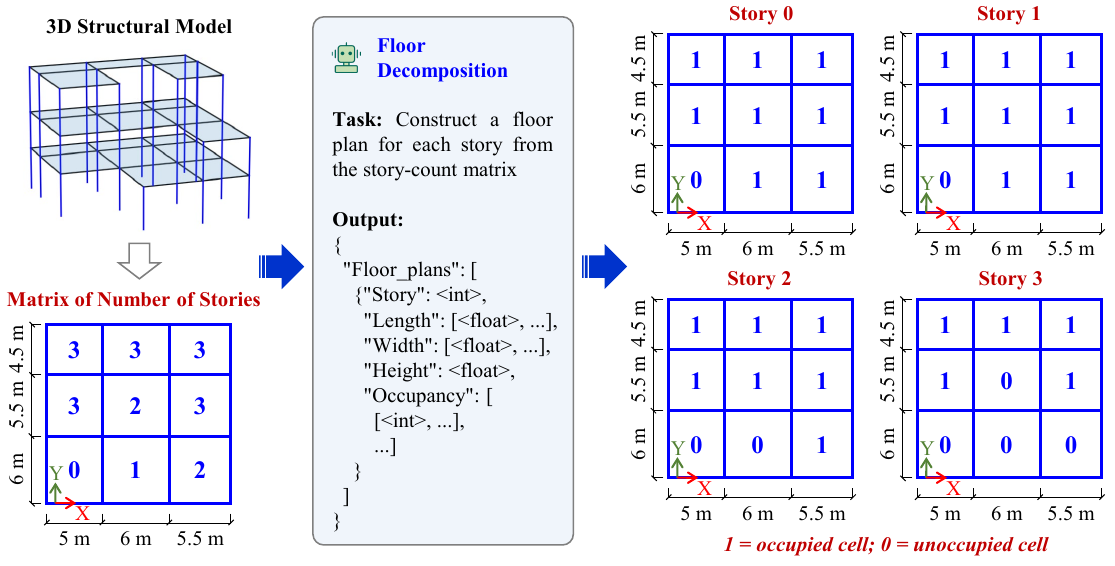}
\caption{Floor decomposition agent for converting the matrix of number of stories to stacked 2D floor plans.}
\label{Figure4}
\end{figure*}

The floor decomposition agent converts the MNS into a set of floor-level occupancy layouts, as illustrated in \cref{Figure4}. For each story, the agent iterates over all grid cells in the matrix and assigns a binary occupancy value: a cell is marked as occupied (1) if its story count is greater than or equal to the current story index, and unoccupied (0) otherwise. For example, a cell with a story count of 3 remains occupied from the first to the third story, whereas a cell with a story count of 1 is active only at the first story. This rule is applied across all cells and story levels, producing a stack of 2D floor plans that collectively encode the 3D topology of the structural system. Each floor plan is represented as a JSON schema containing the story index, gridline coordinates, story height, and occupancy array. This layered representation reduces the burden of single-step geometry reasoning, thereby improving the tractability of geometric generation and mitigating the risk of topological inconsistency.

\begin{figure*}[htbp]
\centering
% \captionsetup{justification=centering}
\includegraphics[width=0.85\textwidth]{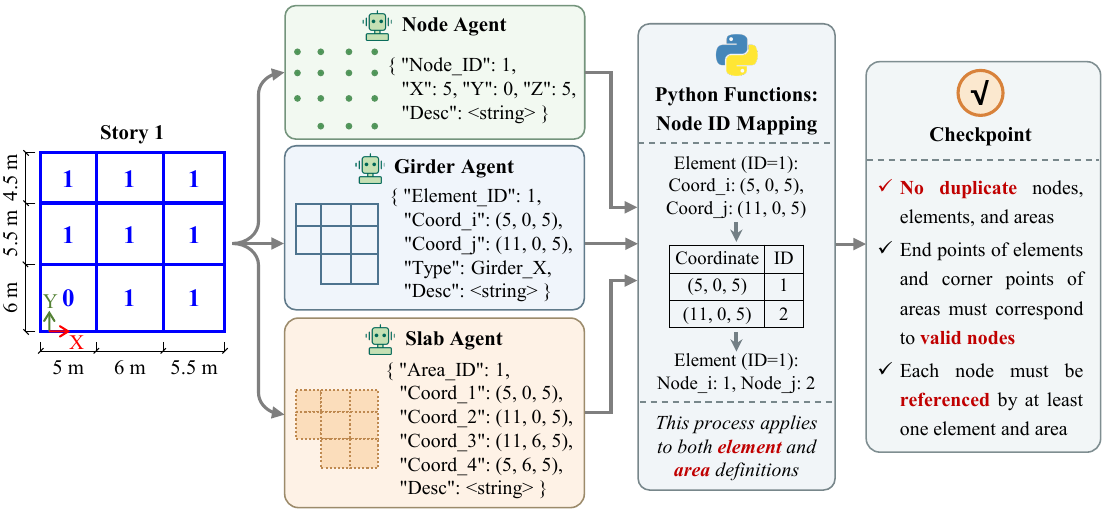}
\caption{Parallel node, girder, and slab agents for in-plane geometry generation.}
\label{Figure5}
\end{figure*}

Given the floor-level occupancy layouts, the node, girder, and slab agents operate in parallel within each floor plan to construct the in-plane structural components, as illustrated in \cref{Figure5}. The node agent derives nodal identifiers and 3D spatial coordinates at gridline intersections associated with occupied cells. The girder agent defines element identifiers, element types, and endpoint coordinates for horizontal members connecting adjacent nodes within the floor plan. The slab agent defines area elements by assigning area identifiers and four corner coordinates to occupied grid cells. All three agents produce structured JSON outputs using a coordinate-based referencing scheme. A Python-based mapping function then converts these coordinate references into corresponding node identifiers required by modeling commands in SAP2000. Following this mapping process, a checkpoint validates the geometric consistency of the generated floor model against three criteria: (i) no duplicate nodes, elements, or areas exist; (ii) all girder endpoints and slab corner points reference valid node identifiers; (iii) each node is connected to at least one element or area. If any criterion is violated, the in-plane geometry is regenerated and revalidated before the workflow continues.

\begin{figure*}[htbp]
\centering
% \captionsetup{justification=centering}
\includegraphics[width=0.85\textwidth]{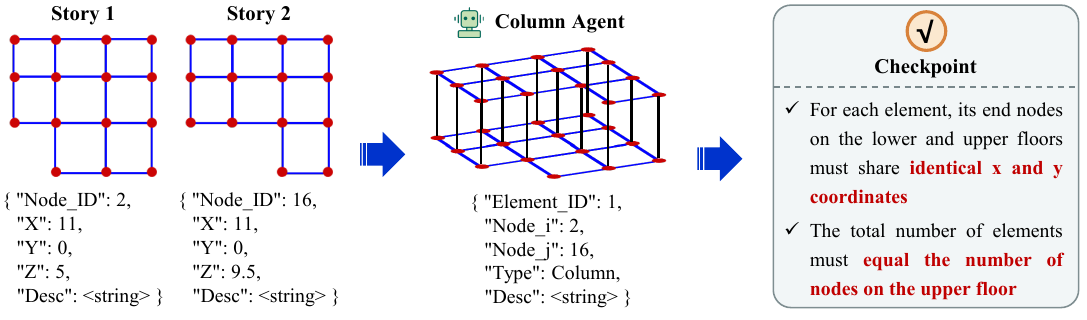}
\caption{Column agent for establishing inter-story connectivity between adjacent floors.}
\label{Figure6}
\end{figure*}

Once the floor-level geometry has been generated and validated, the column agent establishes vertical connectivity between adjacent stories, as shown in \cref{Figure6}. The agent compares the node lists of two consecutive floors and identifies node pairs with identical X–Y coordinates. Each valid pair is connected by a vertical column element. To improve inference efficiency, this process is performed in parallel for all adjacent floor pairs. After column generation, a checkpoint is introduced to verify inter-story topological consistency. For each column, the two end nodes must share the same plan coordinates and differ only in elevation. In addition, the total number of generated columns must be consistent with the number of nodes on the upper floor. Upon successful validation, the complete geometric model, comprising nodal coordinates, in-plane girders, slab areas, and inter-story columns, is compiled into a structured JSON file and passed to the subsequent stages.

\subsection{Support and loading assignment}
\label{sec:stage2}

\begin{figure*}[htbp]
\centering
% \captionsetup{justification=centering}
\includegraphics[width=0.85\textwidth]{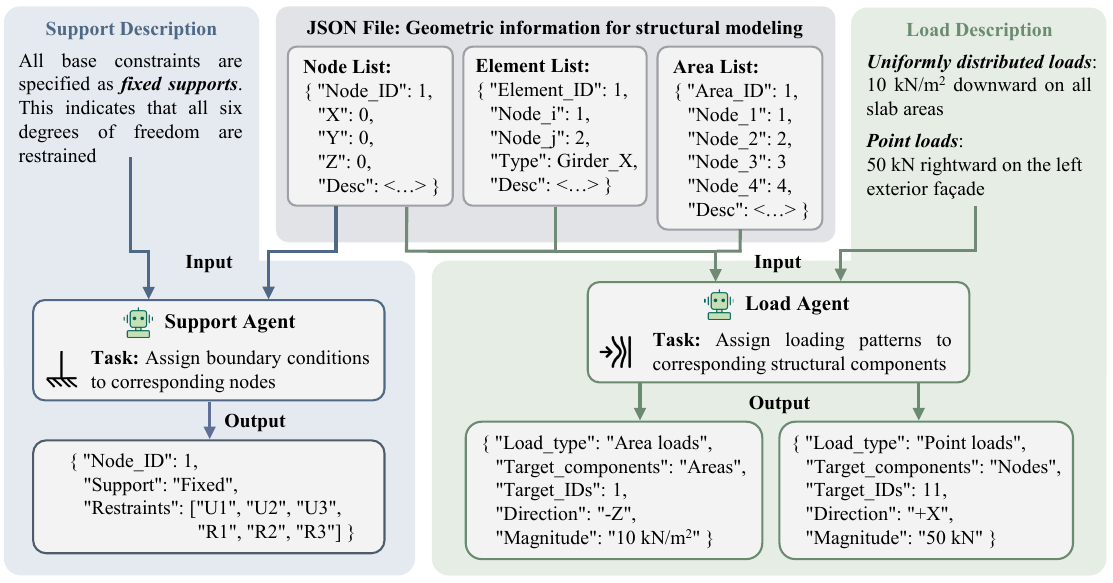}
\caption{Support and load agents for assigning boundary and loading conditions to corresponding structural components.}
\label{Figure7}
\end{figure*}

Following geometry generation, the support and load agents assign boundary conditions and loading patterns to the corresponding structural components, as illustrated in \cref{Figure7}. Both agents receive two types of input: the geometric JSON file generated in Section 3.2 and the support or load descriptions extracted by the problem analysis agent. Specifically, the support agent identifies relevant nodes from the node list based on their spatial coordinates and assigns the corresponding boundary constraints. For example, in the benchmark problems, all supports are defined as fixed at the base. Accordingly, the support agent identifies all nodes located at the base level and restrains all six degrees of freedom, including translational degrees U1, U2, and U3, and rotational degrees R1, R2, and R3. The support agent outputs a structured JSON object in which each constrained node is associated with a support type and its corresponding restrained degrees of freedom.

The load agent maps textual load descriptions to the corresponding structural components using the node, element, and area lists. Herein, two representative load types are used as illustrative examples: uniformly distributed area loads and lateral point loads. For area loads, the agent assigns the specified magnitude and direction to all slab areas in the area list. For point loads, the agent identifies nodes on the left exterior façade and assigns the specified lateral force to the top node of each story. The output of the load agent is a structured JSON object in which each load entry specifies the load type, target component category, target identifiers, direction, and magnitude. The resulting support and load JSON objects are then passed to the code translation stage, where they are converted into modeling commands in SAP2000.

\begin{figure*}[htbp]
\centering
% \captionsetup{justification=centering}
\includegraphics[width=0.85\textwidth]{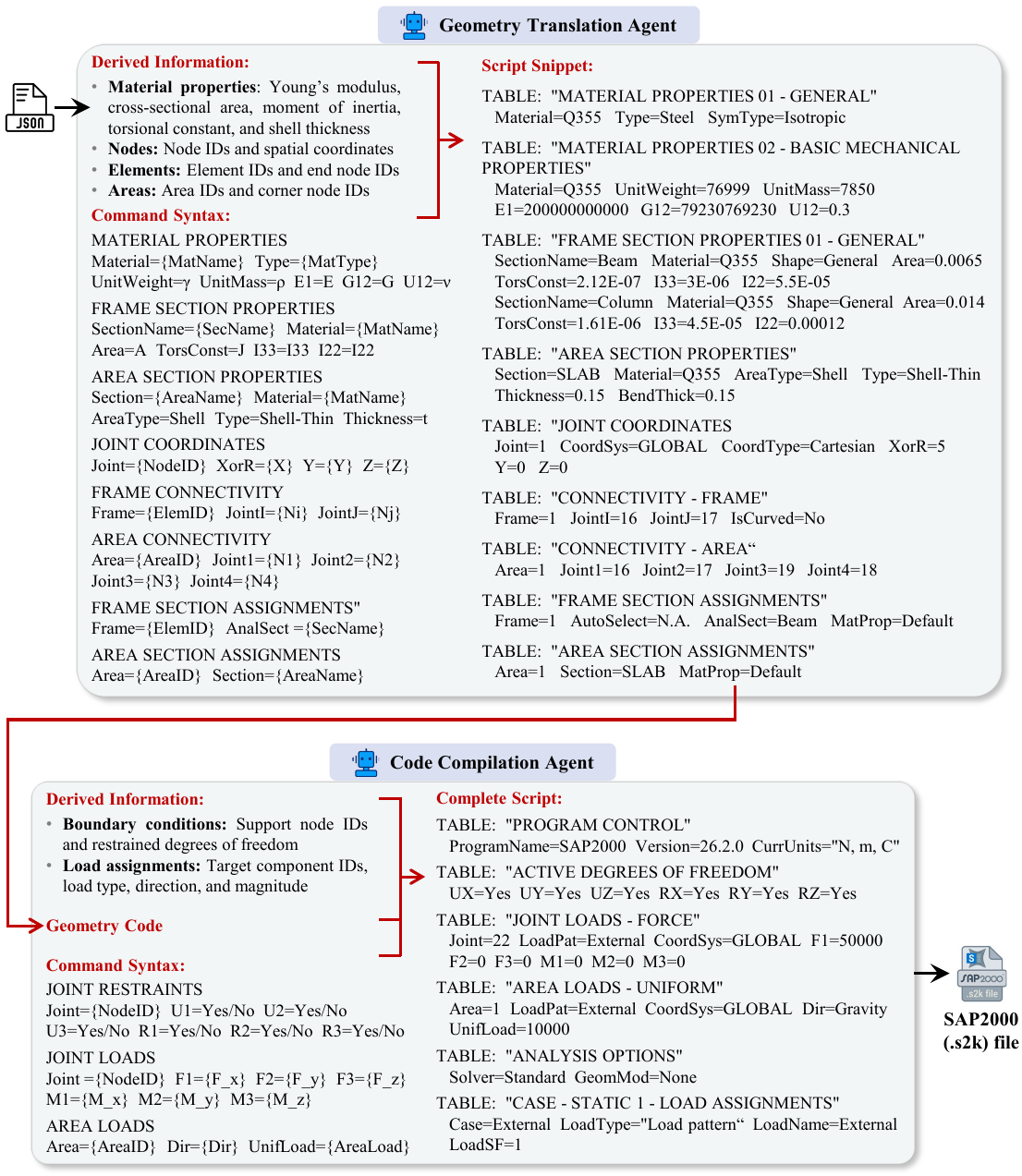}
\caption{Code translation stage for generating executable SAP2000 scripts through geometry translation agent and code compilation agent.}
\label{Figure8}
\end{figure*}

\subsection{SAP2000 script translation}
\label{sec:stage3}

The code translation stage converts all derived modeling information into an executable SAP2000 script through two sequential agents: a geometry translation agent and a code compilation agent, as illustrated in \cref{Figure8}. The geometry translation agent receives two types of input: the geometric JSON file generated in Section 3.2 and the material parameters extracted by the problem analysis agent. Specifically, the material parameters are used to define section properties for frame members and slab areas. The node list is translated into joint definition commands in SAP2000, where each node is defined by its identifier and spatial coordinates. Frame elements and slab areas are defined using their respective identifiers and corresponding node references to establish connectivity. Section assignment commands are subsequently generated to associate each structural component with its designated cross-sectional properties. To ensure syntactic correctness and formatting consistency, a template of SAP2000 command syntax is embedded in the system prompt of the agent.

After the geometric script snippets are generated, the code compilation agent integrates all modules into a complete and executable SAP2000 script. The agent receives the structured outputs of the support and load agents to generate boundary and loading commands. Specifically, boundary conditions are translated into restraint assignment commands that specify the restrained degrees of freedom for each support node, while loading information is converted into load assignment commands that define the target identifiers, load directions, and magnitudes. The code compilation agent also receives the geometry code as input to concatenate it with the boundary and load modules. In addition, configuration blocks such as program control, active degrees of freedom, and analysis options are included in the system prompt to ensure proper model initialization and execution. This two-stage translation strategy mitigates the long-horizon code generation errors and improves the traceability of the agentic workflow.

\section{Results and Discussion}

\subsection{Performance of the proposed agentic LLMs}

\begin{figure*}[htbp]
\centering
% \captionsetup{justification=centering}
\includegraphics[width=0.85\textwidth]{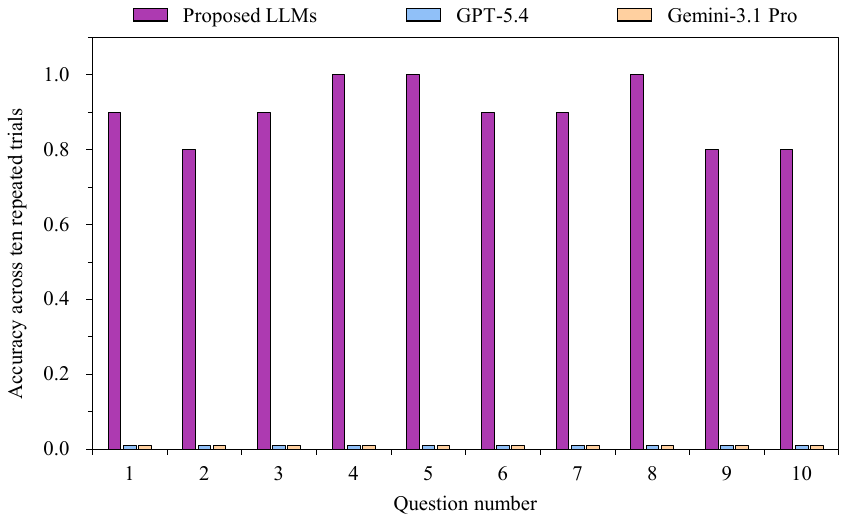}
\caption{Performance comparison between the proposed agentic LLMs and state-of-the-art general-purpose LLMs on the benchmark dataset.}
\label{Figure9}
\end{figure*}

The proposed agentic LLMs are evaluated using the ten representative benchmark problems described in Section 2.2. Each problem is tested over ten repeated trials under identical input conditions. \cref{Figure9} presents the accuracy results across all benchmark cases. It shows that the proposed framework demonstrates consistently strong performance in automated structural analysis of 3D frame systems, achieving accuracy exceeding 80\% for all benchmark problems. Across the ten problems, the average accuracy reaches 90\%, with a variance of 0.007. These results demonstrate that the proposed framework can reliably transform natural language descriptions into geometrically consistent and syntactically executable structural models across a diverse range of irregular 3D frame configurations.

A detailed error analysis is conducted to examine the failure cases of the proposed framework. Two primary error types are identified. The first occurs during section assignment, where frame elements or slab areas are assigned incorrect material properties. For example, beam section properties may be mistakenly assigned to column elements, or section assignments may be omitted for certain areas. These errors can be attributed to the long input context of the geometry translation stage, where the agent must simultaneously process geometric JSON and material properties and map each property to the correct structural component. This places considerable demand on the long-horizon reasoning capacity of lightweight LLMs. The second error type is associated with load assignment. In failed trials, the load agent either omits specified point or area loads at certain locations, or generates additional loads not present in the problem description. This suggests that semantic information about target components may be partially lost when the agent reasons over long geometric contexts. Both failure modes indicate the challenge of long-horizon reasoning in automated structural modeling and highlight the significance of task decomposition and orchestration in the proposed agentic LLMs.

\begin{figure*}[htbp]
\centering
% \captionsetup{justification=centering}
\includegraphics[width=0.85\textwidth]{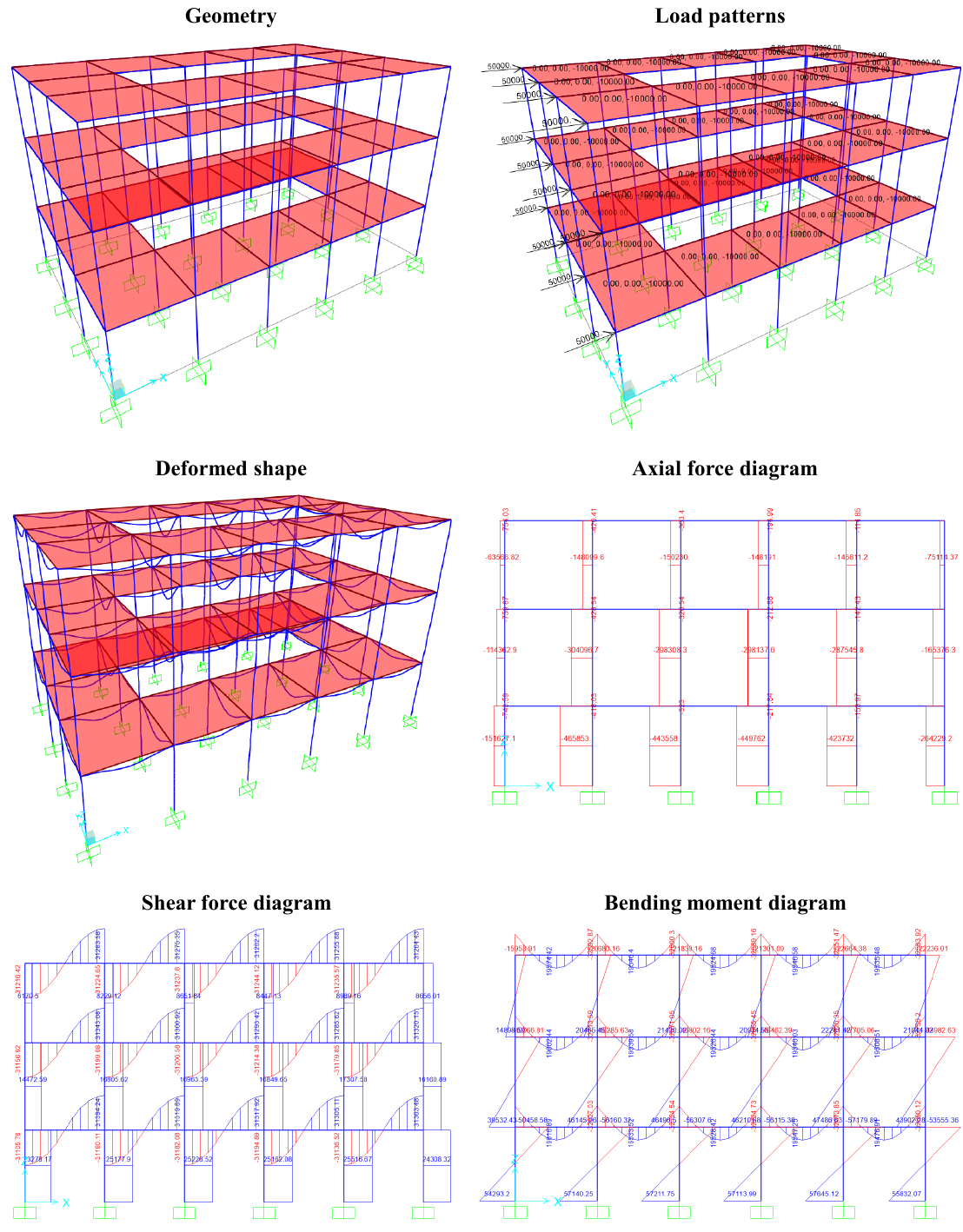}
\caption{Visualization of the generated structural model and analysis results in SAP2000.}
\label{Figure10}
\end{figure*}

The scripts generated by the proposed framework can be directly imported into SAP2000 for structural analysis. This step requires minimal manual intervention because all model configurations have been defined in the scripts. As illustrated in \cref{Figure10}, SAP2000 provides multiple visualization modes that support model inspection and verification. Geometric visualization displays nodal coordinates, frame elements, slab areas, and boundary conditions, allowing users to confirm that the structural geometry has been correctly assembled. Load visualization supports verification of load magnitudes, directions, and target components against the original problem description. These visual checks provide a human-in-the-loop interface for confirming and, if necessary, correcting the generated model before analysis. Upon verification, structural analysis can be performed within SAP2000, yielding mechanical response outputs including deformed shapes, axial force, shear force, and bending moment diagrams. These results can be used for subsequent structural design and engineering decision-making.

\subsection{Comparison with state-of-the-art LLMs}

To further examine the effectiveness of the proposed agentic LLMs, its performance is compared with two SOTA general-purpose LLMs, including GPT-5.4 and Gemini-3.1 Pro. Both baseline models are provided with the same input used by the proposed framework, i.e., the problem description template introduced in Section 2.1, and are instructed to generate complete SAP2000 scripts for structural analysis. As shown in \cref{Figure9}, both models achieve 0\% accuracy across all ten benchmark problems, failing to produce a correct structural model in any repeated trial. These results indicate that, despite their strong general reasoning and code generation capabilities, leading general-purpose LLMs remain insufficient for complex domain-specific tasks such as automated 3D structural modeling and analysis.

A detailed failure analysis is conducted to diagnose the sources of error in the baseline models. The results show that multiple types of errors are observed during SAP2000 script generation. The first barrier occurs at the import stage, where SAP2000 parses the script by exactly matching predefined table names. Any incomplete, misspelled, or nonstandard table name can prevent the script from being imported correctly. GPT model demonstrates a severe unfamiliarity with this domain-specific syntax, with only 4\% of generated scripts being successfully imported on average across the ten cases. Gemini model shows a relative improvement, with an average import success rate of 0.47, but it remains highly unreliable for automated workflows. Even among scripts that are successfully imported, both baseline models frequently generate SAP2000 commands with syntax errors that prevent model execution. Representative errors include incorrect field names in node coordinate definitions, invalid naming conventions for frame elements, missing or inconsistent load pattern definitions, and improper load assignment syntax. These findings indicate that general-purpose LLMs struggle to generate scripts that satisfy the strict syntax requirements of engineering software.

\begin{figure*}[htbp]
\centering
% \captionsetup{justification=centering}
\includegraphics[width=0.85\textwidth]{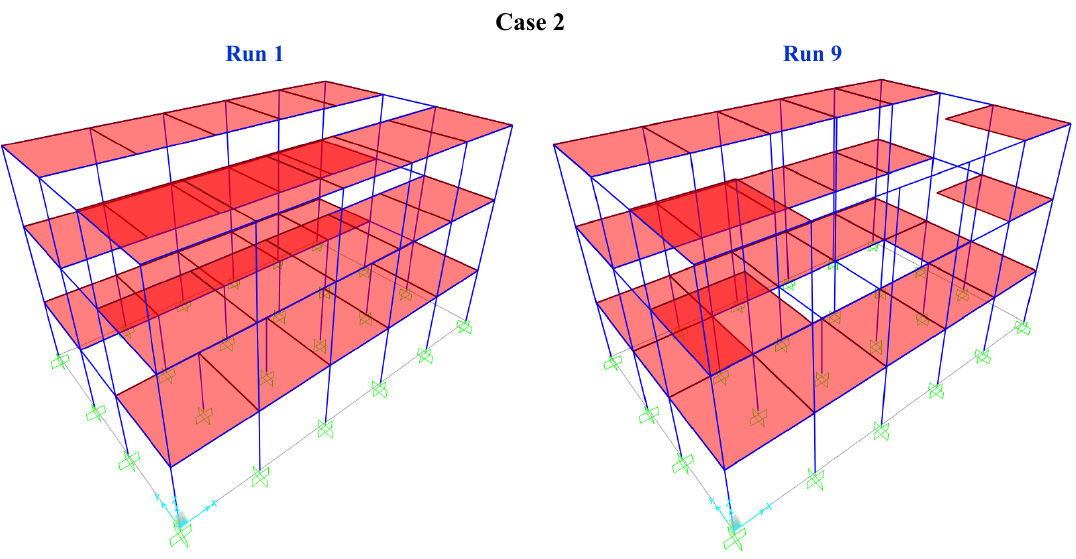}
\caption{Illustrative examples of inconsistent structural geometry generated by GPT-5.4.}
\label{Figure11}
\end{figure*}

\begin{figure*}[htbp]
\centering
% \captionsetup{justification=centering}
\includegraphics[width=0.85\textwidth]{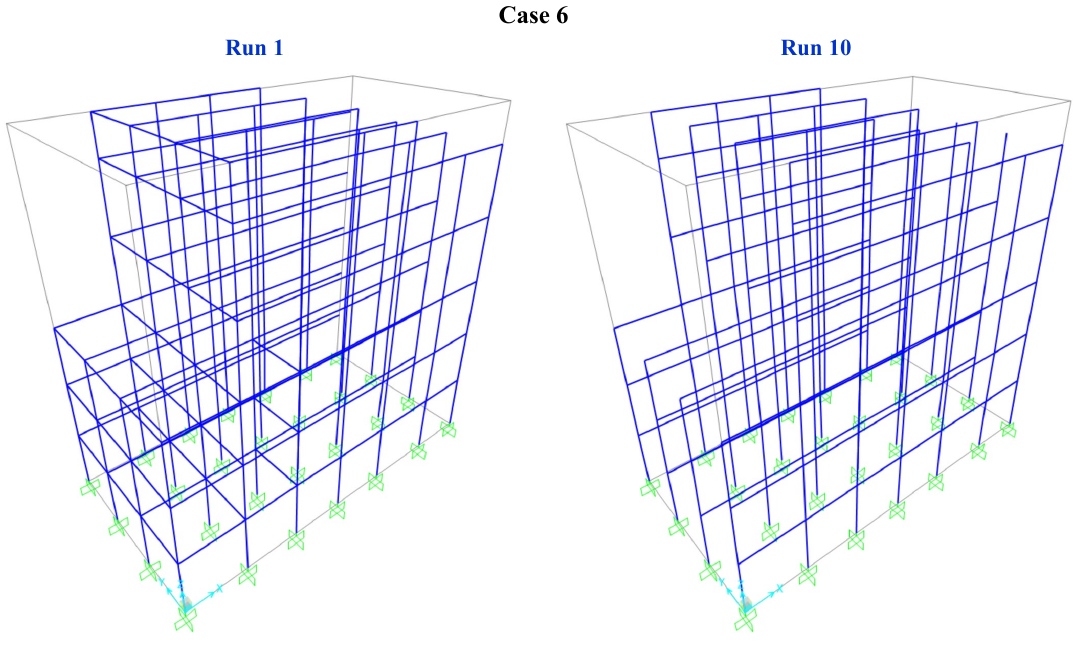}
\caption{Illustrative examples of inconsistent structural geometry generated by Gemini-3.1 Pro.}
\label{Figure12}
\end{figure*}

Beyond syntax errors, GPT and Gemini models fail to maintain geometric and topological consistency across repeated runs. As shown in \cref{Figure11}, GPT model fails to correctly reproduce the open atrium configuration in case 2 and generates inconsistent geometries across trials. In Run 1, the model misinterprets the void layout, extending the atrium region to the full width of the plan. It treats all five bays in the middle row as voids rather than the designated three. In Run 9, the atrium is partially recognized, but slab areas are inconsistently placed, present in some bays and missing in others. Similarly, \cref{Figure12} shows significant variability in the frame connectivity generated by Gemini. In both Run 1 and Run 10, numerous girder elements are omitted across floor plans, compromising the structural integrity of the generated model. These illustrative examples highlight the limitations of general-purpose LLMs in constructing structural topology for irregular 3D frames.

\subsection{Ablation experiments}

To validate the necessity of task decomposition in the proposed agentic LLMs, ablation experiments are conducted by removing or merging specific modules. Specifically, three variants are designed: (i) without the floor decomposition agent, in which the MNS is passed directly from the problem analysis agent to the downstream geometry agents. The geometry agents are instructed to generate nodes, girders, and slabs for the entire 3D frame in a single inference pass; (ii) a merged in-plane geometry agent, in which the node, girder, and slab agents are combined into one agent that defines all in-plane structural components simultaneously; (iii) a merged script translation agent, in which the geometry translation and code compilation agents are combined into one agent to produce the complete SAP2000 script. These variants are evaluated on the first three benchmark cases, with each case repeated ten times under the same input conditions. The results are summarized in Table~\ref{tab:ablation_results}.

\begin{table}[htbp]
\centering
\captionsetup{skip=5pt}
\caption{Ablation experiment results comparing the proposed agentic LLMs with three variants on representative cases.}
\label{tab:ablation_results}
\renewcommand{\arraystretch}{1.15}
\small
\begin{tabular}{
>{\raggedright\arraybackslash}m{4.5cm}
>{\raggedright\arraybackslash}m{5.0cm}
>{\centering\arraybackslash}m{1.0cm}
>{\centering\arraybackslash}m{1.0cm}
>{\centering\arraybackslash}m{1.0cm}
}
\toprule
\textbf{Model} & \textbf{Meaning} & \multicolumn{3}{c}{\textbf{Accuracy}} \\
\cmidrule(lr){3-5}
& & \textbf{Case 1} & \textbf{Case 2} & \textbf{Case 3} \\
\midrule
Proposed agentic LLMs 
& Full framework 
& \textbf{90\%} & \textbf{80\%} & \textbf{90\%} \\

w/o floor decomposition agent 
& Floor decomposition agent removed 
& 30\% & 50\% & 20\% \\

Merged in-plane geometry agent 
& Node, girder, and slab agents merged 
& 50\% & 70\% & 60\% \\

Merged script translation agent 
& Translation and compilation merged 
& 0\% & 0\% & 0\% \\
\bottomrule
\end{tabular}
\end{table}

The removal of floor decomposition agent leads to substantial performance degradation, reducing the accuracy to 30\%, 50\%, and 20\% for cases 1, 2, and 3, respectively. Without floor-level occupancy layouts as intermediate representations, errors frequently arise at the geometric checkpoint, including duplicated nodes, elements with undefined end nodes, and slab areas with invalid corner nodes. This degradation indicates that direct reasoning over the full 3D topology exceeds the reliable reasoning capacity of lightweight LLMs. On the other hand, the merged in-plane geometry agent achieves accuracy of 50\%, 70\%, and 60\% across the three cases. Although this performance exceeds that of the variant without the floor decomposition agent, it remains consistently lower than that of the proposed framework. This result suggests that, even when floor-level layouts are available, the simultaneous generation of nodes, girders, and slabs within a single agent introduces excessive task complexity and weakens topological consistency.

The most severe performance degradation is observed when the geometry translation and code compilation agents are merged into a single script translation agent. This results in a consistent accuracy of 0\% across all three cases. This failure is mainly associated with incorrect material and section assignments, omitted load assignments, and inconsistencies between geometric definitions and subsequent SAP2000 command blocks. Because a complete SAP2000 script for 3D frame modeling can contain more than one thousand lines, these results confirm that code translation is a long-horizon generation task that cannot be reliably handled by a single lightweight LLM without explicit decomposition. Collectively, the ablation results demonstrate that each decomposition strategy in the proposed framework makes a distinct contribution to overall accuracy.

\subsection{Runtime and costs}

\begin{figure*}[htbp]
\centering
% \captionsetup{justification=centering}
\includegraphics[width=0.45\textwidth]{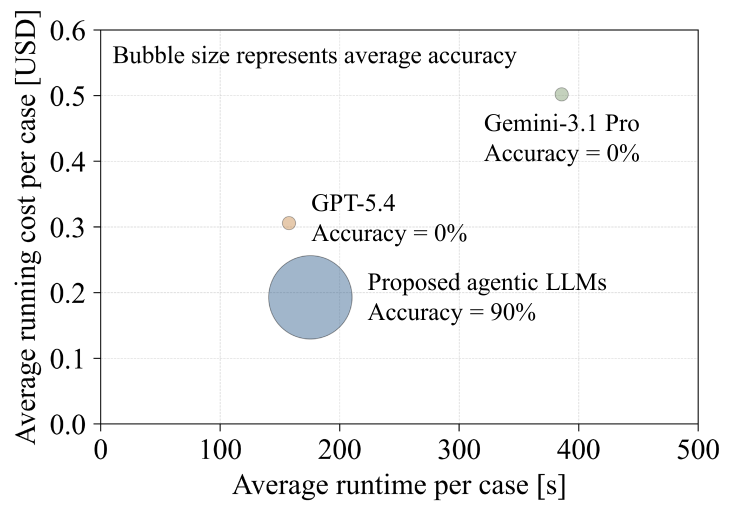}
\caption{Runtime and cost comparison between the proposed agentic LLMs and state-of-the-art general-purpose LLMs.}
\label{Figure13}
\end{figure*}

In addition to accuracy, the runtime and cost of the proposed agentic LLMs are compared with those of the two baseline models across all ten benchmark problems, as shown in \cref{Figure13}. It shows that the proposed framework achieves competitive computational efficiency, with an average runtime of approximately 175 seconds per case. This runtime is comparable to that of the GPT-5.4 model, which requires approximately 158 seconds per case. However, GPT-5.4 fails to generate correct structural models across all cases, rendering its processing speed practically irrelevant. In contrast, the proposed framework is significantly more efficient than Gemini-3.1 Pro, which takes approximately 386 seconds per case on average. This efficiency improvement can be attributed to two architectural design choices: the decomposition of the overall modeling task into lightweight subtasks with narrower reasoning scopes, and the parallel execution of node, girder, and slab agents during geometry generation.

The cost effectiveness of the proposed agentic LLMs is also shown in \cref{Figure13}. The running cost is calculated by multiplying the number of input and output tokens consumed in each API call by the corresponding token price of the selected model, and then aggregating the costs across all agents. The proposed framework incurs an average running cost of USD 0.193 per case, compared with USD 0.306 for GPT-5.4 and USD 0.502 for Gemini-3.1 Pro. This cost advantage is primarily attributable to the use of lightweight open-source LLM backbones, namely GPT-OSS 120B and Llama-3.3 70B Instruct Turbo, whose token pricing is substantially lower than that of frontier commercial models. Although the proposed framework invokes multiple agents, the overall cost remains low because each agent is instructed to produce concise structured outputs, such as JSON objects or SAP2000 scripts. This output design avoids unnecessary explanatory text and reduces token consumption during intermediate reasoning and generation. Collectively, these results demonstrate that the proposed agentic LLMs achieve a compelling balance among accuracy, efficiency, and cost-effectiveness, establishing it as a practically viable solution for automated 3D structural modeling in real-world engineering workflows.

\section{Conclusions and Future Work}

This paper proposes an agentic large language models (LLMs) framework for automated structural analysis of 3D frame systems from natural language inputs. The framework is designed to address three fundamental challenges in this domain: ambiguous representation of irregular geometries, difficulty in maintaining topological consistency, and error accumulation during long-horizon SAP2000 script generation. To represent irregular 3D frames, a structured geometric description scheme is introduced, where the 3D frame is projected onto a 2D plan defined by orthogonal gridlines, while a matrix of number of stories (MNS) specifies the vertical extrusion within each grid cell. Building on this representation, the proposed framework decomposes the overall modeling process into a sequence of subtasks, each delegated to a specialized agent. Specifically, a problem analysis agent extracts geometric, boundary, loading, and material information from the textual input and encodes them into a structured JSON schema. A floor decomposition agent then converts the MNS into floor-level occupancy layouts. Based on these layouts, node, girder, and slab agents operate in parallel to generate in-plane structural components, while a column agent establishes vertical connectivity between adjacent stories. Support and load agents assign boundary conditions and loading patterns to the corresponding nodes, elements, and slab areas. Finally, a geometry translation agent and a code compilation agent convert the structured information into an executable SAP2000 script. Checkpoints are embedded throughout the workflow to detect errors before information is passed to downstream stages, improving the geometric and topological consistency of the generated models. The proposed framework is evaluated on ten representative 3D frame problems spanning a diverse range of irregular geometric configurations. The key findings are summarized below:

\begin{itemize}[leftmargin=1.5em, itemsep=4pt, topsep=3pt]

    \item The proposed agentic LLMs demonstrate consistently high and reliable performance for automated 3D structural modeling. Across ten benchmark problems, the framework achieves an average accuracy of 90\% with low variance over ten repeated trials per case. The SAP2000 script generated can be directly imported and executed, producing structural responses that closely match those of manually constructed ground truth models.

    \item The proposed framework significantly outperforms state-of-the-art (SOTA) general-purpose LLMs. GPT-5.4 and Gemini-3.1 Pro fail to generate correct structural models across all benchmark cases when prompted to generate SAP2000 scripts directly from natural language inputs. Failure analysis shows that these baseline models suffer from SAP2000 syntax errors and severe geometric and topological inconsistencies.
    
    \item Ablation experiments confirm that each task decomposition strategy contributes distinctly to overall performance. Removing the floor decomposition agent reduces accuracy to 30–50\%; merging the node, element, and slab agents reduces accuracy to 50–70\%; and consolidating the code translation stage into a single agent results in complete failure across all tested cases. These results validate the necessity of structured task decomposition for reliable long-horizon structural modeling.
    
    \item The current framework is limited to 3D frame systems whose floor plans can be discretized into rectangular grid cells. It cannot yet represent nonorthogonal or curvilinear geometries, which may be difficult to describe using natural language alone. Future work should integrate vision language models (VLMs) to incorporate visual information and extend the framework to a broader class of irregular structural geometries.

    \item The current framework is restricted to static structural analysis and does not support lateral force resisting systems. Future work will extend the framework to encompass dynamic analysis capabilities, including seismic and wind-induced response simulation. Also, it will incorporate shear wall and bracing modeling into the multi-agent pipeline to improve the applicability to realistic structural design and analysis scenarios. 
    
\end{itemize}

\section*{Data Availability Statement}
Some or all data, models, or code that support the findings of this study are available from the corresponding author upon reasonable request. 

\section*{Acknowledgments}
The authors are grateful for the financial support received from US Department of Transportation Tier 1 University Transportation Center CREATE Award No. 69A3552348330.

\bibliographystyle{ascelike}  
\bibliography{references}
%%% Remove comment to use the external .bib file (using bibtex).


% George Kour and Raid Saabne.
% \newblock Real-time segmentation of on-line handwritten arabic script.
% \newblock In {\em Frontiers in Handwriting Recognition (ICFHR), 2014 14th
%   International Conference on}, pages 417--422. IEEE, 2014.

% \end{thebibliography}

\end{document}